\documentclass[preprint,aps,prc,tightenlines,
floatfix,showpaces,showkeys]{revtex4}

\usepackage{graphicx}
\usepackage{bm}
\usepackage{amsmath}
\usepackage{amssymb}

\def\ss{\mbox{\boldmath $\sigma$}}
\def\tt{\mbox{\boldmath $\tau$}}
\def\bfrho{\mbox{\boldmath $\varrho$}}
\def\bfchi{\mbox{\boldmath $\chi$}}
\newcommand{\be}{\begin{equation}}
\newcommand{\ee}{\end{equation}}
\newcommand{\bea}{\begin{eqnarray}}
\newcommand{\eea}{\end{eqnarray}}
\newcommand{\nn}{\nonumber}

\newcommand{\Image}{\mathop{\rm Im}\nolimits}

\newcommand{ \bb }{$2\nu\beta\beta$}
\newcommand{ \bbm }{$2\nu\beta^-\beta^-$}
\newcommand{ \bbzn }{$0\nu\beta\beta$}

\begin{document}

\title{ \vspace{1cm} Description of double beta decay within continuum-QRPA}
\author{Vadim Rodin}
\email{vadim.rodin@uni-tuebingen.de}
\author{Amand Faessler}
\email{amand.faessler@uni-tuebingen.de}
\affiliation{Institute for Theoretical Physics, University of T\"ubingen, Auf der Morgenstelle 14, D-72076 T\"ubingen, Germany}

\date{\today}

\begin{abstract} 
A method to calculate the nuclear double beta decay (\bb \ and \bbzn) amplitudes 
within the continuum quasiparticle random phase approximation (cQRPA) is formulated.
Calculations of the $\beta\beta$ transition amplitudes 
within the cQRPA are performed for $^{76}$Ge, $^{100}$Mo and $^{130}$Te.
A rather simple nuclear Hamiltonian consisting of a phenomenological mean field 
and a zero-range residual particle-hole and particle-particle interaction is used.
The calculated \bb \ amplitudes 
are almost unaffected when the single-particle continuum is taken into account, whereas we find 
a regular suppression of the \bbzn \ amplitudes 
that can be associated with additional ground-state correlations 
owing to collective states in the continuum. It is expected that inclusion
of nucleon pairing in the single-particle continuum will somewhat compensate this suppression.

\pacs{
21.60.-n, 
21.60.Jz, 
23.40.-s, 
23.40.Hc, 
}

\keywords{Double beta decay; Nuclear matrix element; Quasiparticle random phase approximation}
\end{abstract}

\maketitle

\section{Introduction}
Neutrino oscillation experiments have proven that neutrinos are massive particles (see, e.g., Ref.~\cite{McK04}). 
However, the absolute scale of the neutrino masses cannot in principle be deduced from the observed oscillations.
To determine the absolute neutrino masses down to the level of tens of meV,
study of the neutrinoless double beta decay ($0\nu\beta\beta$) becomes indispensable.
Furthermore, this process, which violates the total lepton number by two units, is an
{\em experimentum crucis} to reveal the Majorana nature of neutrinos~\cite{fae98,Suh98,EV2002,El04}.

The next generation of experiments (GERDA, CUORE, SuperNEMO etc.)
has a great discovery potential for observation of $0\nu\beta\beta$ \ decay 
and for providing reliable measurements of the corresponding lifetimes. 
The determination of the effective Majorana mass (or relevant GUT and SUSY parameters depending on 
what mechanism of the $0\nu\beta\beta$ \ decay dominates)
from experimental data can be only as good as the knowledge of the nuclear matrix elements $M^{0\nu}$ 
on which the $0\nu\beta\beta$\ decay rates depend. 
Thus, a better understanding of the nuclear structure effects important for describing the matrix
elements is needed to interpret the data accurately. 
It is crucial in this connection to develop theoretical methods capable of reliably evaluating 
the nuclear matrix elements, and to realistically assess their uncertainties.

At present, the most elaborate analysis 
of the uncertainties in the \bbzn \ decay nuclear matrix elements $M^{0\nu}$ calculated within the quasiparticle random phase approximation (QRPA) and the renormalized quasiparticle random phase approximation (RQRPA) has been performed in recent papers~\cite{Rod03a,Rod05} (the bases comprising $N=$2, 3 and 5 major oscillator shells were used). The experimental \bbm \ decay rates 
have been used there to adjust the most relevant parameter, the strength $g_{pp}$ of the particle-particle
interaction. The major observation of Refs.~\cite{Rod03a,Rod05} is that 
such a procedure makes the calculated $M^{0\nu}$ essentially independent of the size of the single-particle (s.p.) basis of the QRPA . Furthermore, the matrix elements 
have been demonstrated to also become rather stable with respect  
to the possible quenching of the axial vector coupling constant $g_A$.

The calculations in Refs.~\cite{Rod03a,Rod05} were performed within ``the standard QRPA" scheme in which 
a discrete s.p. basis and the harmonic oscillator wave functions as the s.p. wave functions are used to build the BCS ground state and the spectrum of the excited states.
Keeping in mind that many multipoles contribute appreciably to $M^{0\nu}$, one can {\it a priori} expect that enlargement of the model space should lead to more accurate matrix elements $M^{0\nu}$. (In other words, any basis truncation leads to an uncertainty.) This should be contrasted with the case of the \bb \ amplitude to which only Gamow-Teller transitions contribute and 
a s.p. basis of $N=$1-2 major shells is good enough. 
In this respect, it would be interesting to test the stability of the calculated $M^{0\nu}$ found in Refs.~\cite{Rod03a,Rod05} by letting $N\to\infty$. Thus, if one could include the entire s.p. basis into the calculation scheme, 
the question about the dependence of the QRPA results on size of the s.p. basis as a source of the uncertainties in the calculated $M^{0\nu}$ would become irrelevant.

There is no problem within the QRPA for including low-lying major shells 
composed of bound s.p. states into the model space.
But inclusion of major shells lying much higher than the Fermi level
immediately encounters principal limitations of approximation of the continuum of unbound s.p. states by discrete levels.
Basically, only one major shell, lying higher than the Fermi shell (already containing quasistationary states),
can safely be considered.

The only possible way to treat properly the s.p. continuum
is provided within the continuum quasiparticle random phase approximation (cQRPA). 
The continuum random phase approximation (cRPA) was formulated about 30 years ago
in the pioneering work of Shlomo and Bertsch~\cite{Shlomo75} and since then has been used to successfully describe structure and decay properties of various giant resonances~\cite{Rod}, muon capture~\cite{Ur92muon}, and neutrino-nucleus reactions with large momentum transfer~\cite{Kolb03}.
To apply the cRPA in open-shell nuclei one has to take nucleon pairing into consideration. This requires development of a quasiparticle version of the cRPA, namely, a continuum QRPA approach. 
Such a cQRPA approach to describing charge-exchange excitations has been developed in Refs.~\cite{Bor90,Rod03}.

The cQRPA provides a regular way of using realistic wave functions of unbound s.p. states
in terms of the s.p. Green's functions without the need to approximate them by the oscillator ones.
Moreover, having an alternative formulation of the QRPA can help to understand current QRPA results and their deficiencies.

Two principal effects of taking into account the s.p. continuum within the proton-neutron QRPA (pn-QRPA), which affect the calculated values of $M^{0\nu}$ in an opposite way, can be expected. 
First, additional ground-state (g.s.) correlations can appear because of collective multipole states in the continuum, 
which generally have a tendency to decrease $M^{0\nu}$. 
Second, pairing in the continuum can increase $0\nu\beta\beta$ matrix element $M^{0\nu}$ 
(see the relevant discussion in Ref.~\cite{Sim07}).

The principal aim of this work is to formulate for the first time a nuclear structure framework
for calculating the double beta decay matrix elements $M^{2\nu}$ and $M^{0\nu}$ within the cQRPA 
and to test within this method the stability of the calculated $M^{0\nu}$ found in Refs.~\cite{Rod03a,Rod05}. 
As a first step, a simpler version of the cQRPA with nucleon pairing realized 
only on a discrete basis is applied in the present work; therefore, the calculated $M^{0\nu}$'s of this paper 
should be considered lower limits for the matrix elements within the cQRPA. 
To consistently include nucleon pairing in the continuum within the cQRPA is a formidable task and is postponed to future publications.

We merely focus here on a qualitative discussion of the relative effect obtained within the cQRPA in comparison with the standard discrete QRPA. Therefore, the $M^{0\nu}$ values of the present work may be somewhat different from those of Refs.~\cite{Rod03a,Rod05} since we have not implemented the most elaborate
representation for the neutrino potential (modified by the finite nucleon size correction, higher order terms of 
the nucleon weak current, etc.; see, e.g.,~\cite{Rod05}). The nucleon-nucleon short-range correlations (SRC) are implemented here in the usual way, in terms of the Jastrow-like functions~\cite{Jastrow}. This, however, might lead to a overestimation of the effect of the SRC (see the recent discussion in Refs.~\cite{Sim07,Suh07a}).

The paper is organized as follows:
The pn-QRPA equations in the coordinate presentation and the way to take into account the s.p. continuum in them 
are given in the first two parts of Sec. II. In the latter two parts of that section formulas for 
calculating strength functions and $M^{2\nu}$ and $M^{0\nu}$ are presented. In Sec.~\ref{results} we present the results and we give conclusions in Sec.~\ref{conclusions}.

\section{Continuum QRPA}

Since its formulation in the pioneering work of Shlomo and Bertsch~\cite{Shlomo75}, the cRPA has long been used for
to successfully describe structure and decay properties of various giant resonances
and their high-lying overtones embedded in the single-particle continuum.
The structure of the overtones 
is formed by the s.p. excitations changing the s.p. radial quantum number (which correspond
to transitions over two or more major shells).
Their contribution to the nuclear multipole response is marked if probe operators have nontrivial radial dependence, 
which is the case, for example, for muon capture~\cite{Ur92muon} and neutrino-nucleus reactions with large momentum transfer~\cite{Kolb03}.  The direct nucleon decay of various giant resonances and their overtones has been extensively analyzed within the cRPA by Urin and collaborators (see, e.g.,~\cite{Rod}).

To apply the cRPA in open-shell nuclei one has to take nucleon pairing into consideration. 
This requires development of a quasiparticle version of the cRPA, namely, the cQRPA approach. 
The approach should account for the important influence of the residual
particle-particle (p-p) interaction along with the particle-hole (p-h) one included within the usual cRPA. 
Such a cQRPA approach based on the coordinate space Hartree-Fock-Bogolyubov formalism
has been formulated and applied recently to describe strength functions of different multipole excitations 
without charge exchange~\cite{Mats01}.

A pn-cQRPA approach to describing charge-exchange excitations was developed in Ref.~\cite{Bor90} 
and, independently, in Ref.~\cite {Rod03}. 
In Ref.~\cite{Bor00} the approach was applied to analyze the low-energy part of the Gamow-Teller (GT) strength distribution  relevant for description of single beta decay in astrophysical applications. 
In ~\cite{Rod03} the Fermi and GT strength distributions in semimagic nuclei were described
within a wide excitation-energy interval that includes the overtones of the IAS and GTR, 
the so-called monopole and spin-monopole resonances.

In describing the \bbzn \ decay, some transition strength into the s.p. continuum is missing within the standard QRPA calculation scheme, especially for the high-multipole excitations with $L\ge 2$ (compare, e.g., with 
the description of muon capture where the contribution of the highly excited giant resonances dominates~\cite{Ur92muon}).
The contribution of these multipoles to $M^{0\nu}$ becomes particularly important
because the monopole (Fermi and Gamow-Teller) contributions are suppressed by symmetry
constraints (see, e.g., the multipole decomposition of $M^{0\nu}$ in Fig.~5 of Ref.~\cite{Rod05};
for a recent general discussion of how the SU(4)-symmetry violation 
by the residual p-p interaction affects $M^{2\nu}$ see Ref.~\cite{Rodin05}).
Thus, $M^{0\nu}$ gets strongly suppressed by the g.s. correlations, short-range correlations, etc.; 
therefore fine effects (such as influence of the s.p. continuum) can be expected to come into play.
The first attempt to briefly describe the $\beta\beta$ observables within the pn-cQRPA was undertaken 
in Ref.~\cite{RodErice}.

\subsection{pn-QRPA equations in coordinate representation \label{QRPA}}

The system of homogeneous equations for the forward and backward amplitudes $X^{(J^\pi s)}_{\pi\nu}$ and $Y^{(J^\pi s)}_{\pi\nu}$,
respectively, is usually solved to calculate the energies $\omega_s$ and the wave functions
$|J^\pi M, s\rangle$ of excited states in isobaric odd-odd nuclei within 
the pn-QRPA (see, e.g.,~\cite{RingSchuck80,fae98}; here ``$s$" labels the different QRPA states). 
However, it is impossible to handle an infinite number of 
amplitudes $X,Y$ if one wants to include the continuum of unbound s.p. states. 
Instead, by going into the coordinate representation 
the pn-QRPA can be reformulated in equivalent terms of four-component radial transition density 
$\{\varrho^{(JLS)}_{I,s}\}\ (I=1,\dots ,4)$ defined for each state $|J^\pi M, s\rangle$. The components are determined by 
the standard QRPA amplitudes $X$ and $Y$ as follows:

\bea
&& \varrho^{(JLS)}_{I,s}(r)=\sum\limits_{\pi\nu} R^{\pi\nu}_{I,s} \, 
\chi_{\pi\nu}(r),\label{defvarrho}\\
&&\left({\begin{array}{c}
R^{\pi\nu}_{p-h}\\
R^{\pi\nu}_{h-p}\\
R^{\pi\nu}_{p-p}\\
R^{\pi\nu}_{h-h}
\end{array}}\right)_s
=
\left({\begin{array}{c}
u_\pi v_\nu X_{\pi\nu}+v_\pi u_\nu Y_{\pi\nu}\\
u_\pi v_\nu Y_{\pi\nu}+v_\pi u_\nu X_{\pi\nu}\\
u_\pi u_\nu X_{\pi\nu}-v_\pi v_\nu Y_{\pi\nu}\\
u_\pi u_\nu Y_{\pi\nu}-v_\pi v_\nu X_{\pi\nu}
\end{array}}\right)_s
\nn
\eea
where $u$ and $v$ are the coefficients of Bogolyubov transformation and 
$\chi_{\pi\nu}(r)=t^{(JLS)}_{(\pi)(\nu)}\,\chi_\pi(r)\chi_\nu(r)$ with $(\pi)=(j_{\pi}l_{\pi})$ [$(\nu)=(j_{\nu}l_{\nu})$] and $r^{-1}\chi_{\pi}(r)$ [$r^{-1}\chi_{\nu}(r)$] being the s.p. proton (neutron) quantum numbers and radial wave functions, respectively.
In Eq.~(\ref{defvarrho}) the spin-angular variables are already separated out 
since the nuclear response to a probe operator having definite spin-angular symmetry 
determined by the irreducible spin-angular tensor $T_{JLSM}({\mathbf n})$ is calculated, 
and $t^{(JLS)}_{(\pi)(\nu)}=\frac{1}{\sqrt{2J+1} }\langle\pi\|T_{JLS}\|\nu\rangle$ represents the corresponding reduced matrix element. Hereafter, we shall systematically omit the superscript ``$(JLS)$" when it does not lead to a confusion.

According to the definition of Eq.~(\ref{defvarrho}), the elements
$\varrho_1,\ \varrho_2,\ \varrho_3$ and $\varrho_4$ can be called 
the particle-hole, hole-particle, particle-particle, and hole-hole
components of the transition density, respectively, and 
can be generally considered as a four-dimensional vector:$\{ \varrho^{J}_I\}$.
In particular, the transition matrix element to a state $|s,JM\rangle$ corresponding
to a particle-hole operator 
\bea
&\hat V^{(-)}_{JLSM}=\sum_a {\mathbf V}_{JLSM}({\mathbf x}_a)\ \tau^-_a      \label{probop}\\
&{\mathbf V}_{JLSM}({\mathbf x}_a)=V_{(JLS)}(r_a)T_{JLSM}({\mathbf n}_a) \label{Vm}
\eea
is determined by a one-dimensional integral of the product of the element $\varrho_1$ and the radial dependence of the operator Eqs.~(\ref{probop},\ref{Vm}):
\be 
\langle J^\pi M|\hat V^{(-)}_{JLSM}|0\rangle=\int \varrho^{(JLS)}_{1}(r) V_{(JLS)}(r)\, dr.
\ee

The pn-QRPA system of integral equations for the elements $\varrho^{(JLS)}_{I,s}$ follows from the standard pn-QRPA equations for the $X$ and $Y$ amplitudes (see, e.g., Refs.~\cite{fae98,RingSchuck80}) by making use of 
the definition from Eq.~(\ref{defvarrho}):
\be
\varrho^{(JLS)}_{I,s}(r)=
\sum\limits_{K} \int A^{(JLS)}_{IK}(rr',\omega=\omega_s)\, F^{(JLS)}_K(r'r'')\, \varrho^{(JLS)}_{K,s}(r'')
\, dr'dr''.\label{eq1varrho}
\ee
Here, 
$(rr')^{-1}F^{(JLS)}_{K}(r,r')$ is the radial part of the residual interaction in the $K$ channel 
(where $K=1,2$ for the p-h channel, $K=3,4$ for the p-p channel and the so-called symmetric approximation is used here).
The $4\times 4$ matrix $(rr')^{-1}A_{IK}(r_1r_2,\omega)$ is the radial part
of the free two-quasiparticle propagator (response function):
\bea
&A_{IK}(r_1r_2,\omega)=\sum\limits_{\pi\nu} 
\, A^{\pi\nu}_{IK}(\omega)\, \chi_{\pi\nu}(r_1)\chi_{\pi\nu}(r_2); \ \ \ \ A^{\pi\nu}_{KI}=A^{\pi\nu}_{IK} 
\label{A}\\
& \nonumber\\
& A^{\pi\nu}_{11}=\displaystyle\frac{u^2_\pi v^2_\nu}{\omega-E_{\pi\nu}} + \frac{u^2_\nu v^2_\pi}{-\omega-E_{\pi\nu}},
\ \ A^{\pi\nu}_{33}=\frac{u^2_\pi u^2_\nu}{\omega-E_{\pi\nu}} + \frac{v^2_\nu v^2_\pi}{-\omega-E_{\pi\nu}},
\nonumber\\
&A^{\pi\nu}_{13}=u_\nu v_\nu \displaystyle\left(\frac{u^2_\pi}{\omega-E_{\pi\nu}} - \frac{v^2_\pi}{-\omega-E_{\pi\nu}}\right),\ \
A^{\pi\nu}_{14}=-u_\pi v_\pi \left(\frac{v^2_\nu}{\omega-E_{\pi\nu}} - \frac{u^2_\nu}{-\omega-E_{\pi\nu}}\right)\ ,\nonumber\\
&A^{\pi\nu}_{12}=-A^{\pi\nu}_{34}=\displaystyle\frac{u_\pi v_\pi v_\nu u_\nu}{\omega-E_{\pi\nu}} + 
\frac{u_\pi v_\pi v_\nu u_\nu}{-\omega-E_{\pi\nu}}\nonumber\\
& A^{\pi\nu}_{22}(\omega)=A^{\pi\nu}_{11}(-\omega),\ \ A^{\pi\nu}_{44}(\omega)=A^{\pi\nu}_{33}(-\omega), \ \ A^{\pi\nu}_{23}(\omega)=A^{\pi\nu}_{14}(-\omega),\ \ A^{\pi\nu}_{24}(\omega)=A^{\pi\nu}_{13}(-\omega).
\nonumber
\eea
with $E_{\pi\nu}=E_{\pi}+E_{\nu}$, where $E_{\pi}$ and $E_{\nu}$ are the proton and neutron
quasiparticle energy, respectively.
The expressions for the elements of the free two-quasiparticle propagator $A_{IK}$
can also be obtained by making use of the regular and anomalous s.p. Green's
functions for Fermi systems with nucleon pairing (see, e.g.,~\cite{Rod03}), in an analogous way to that described 
in the monograph~\cite{Mig83} for response of Fermi systems to a
s.p. probe operator acting in the neutral channel.

These equations allow a compact schematic representation when the spin-angular variables are not separated out.
In such a case, the substitutions $r_a\to \mathbf{x}_a$, $\chi_\alpha(r)\to \bfchi_\alpha(\mathbf{x})$, 
$\varrho\to\bfrho$, $A_{IK}\to\mathbf{A}_{IK}$, and $F_{K}\to\mathbf{F}_{K}$ have to be made 
and the factor $t^{(JLS)}$ should be omitted in the formulas. 
Then, schematically denoting the double integration over $\mathbf{x'},\mathbf{x''}$ in eq.~(\ref{eq1varrho}) as $\{\dots\}$, one can rewrite the equation as 
\be
\bfrho_I=\{\mathbf{A}_{IK}\mathbf{F}_K\bfrho_K\},
\ee
where summation over the repeated index $K$ on the right-hand side is assumed.

The total two-quasiparticle propagator (two-quasiparticle Green function) $\tilde A$ that includes the QRPA iterations of the p-h and p-p interactions is very useful in practical applications. It satisfies an integral equation of the 
Bethe-Salpeter type $\mathbf{\tilde A}_{IK}=\mathbf{A}_{IK}+\{\mathbf{A}_{IK'}\mathbf{F}_{K'}\mathbf{\tilde A}_{K'K} \}$:
\bea
&\tilde A^{(JLS)}_{IK}(rr',\omega)=A^{(JLS)}_{IK}(rr',\omega)+
\sum\limits_{K'} \int A^{(JLS)}_{IK'}(rr_1,\omega) \, F^{(JLS)}_{K'}(r_1r_2)\, \tilde A^{(JLS)}_{K'K}(r_2r',\omega)
\, dr_1dr_2.
\label{Atilde}
\eea

The following spectral decompositions hold for 
the radial components $\tilde A_{11}(\omega),\tilde A_{12}(\omega),$ and $\tilde A_{22}(\omega)$:
\bea
&\tilde A_{11}(r_1r_2,\omega)=\displaystyle
\sum_s \frac{\varrho^s_{1}(r_1)\varrho^s_{1}(r_2)}{\omega-\omega_s+i\delta}
-\sum_s \frac{\varrho^s_{2}(r_1)\varrho^s_{2}(r_2)}{\omega+\omega_s-i\delta}\nonumber\\
&\tilde A _{22}(r_1r_2,\omega)=\tilde A _{11}(r_1r_2,-\omega)\label{spectrA}\\
&\tilde A_{12}(r_1r_2,\omega)=\displaystyle
\sum_s \frac{\varrho^s_{1}(r_1)\varrho^s _{2}(r_2)}{\omega-\omega_s+i\delta}
-\sum_s \frac{\varrho^s_{2}(r_1)\varrho^s_{1}(r_2)}{\omega+\omega_s-i\delta}
\nonumber
\eea

These components are the only ones that will be needed in the following
for description of single and double beta decay transition probabilities. 
The spectral decompositions for the other elements of $\tilde A$ 
can readily be written down in analogy to Eqs.~(\ref{spectrA}).
Thus, one sees from Eqs.~(\ref{spectrA}) that all the information about the QRPA solutions 
[energies $\omega_s$ and transition densities $\varrho_{I,s}(r)$] resides
in the poles of the total two-quasiparticle propagator $\tilde A$.

In this paper, the residual isovector particle-hole interaction 
and the particle-particle interaction in both 
the neutral (pairing) and charge-exchange channels are chosen in the form of the 
Landau-Migdal forces of zero range (proportional to the 
spatial $\delta$ function)~\cite{Mig83}, which is similar to the choice of Refs.~\cite{vog86,Engel88}. 
The effective isovector particle-hole interaction $\mathbf{F}_K$ (for $K=1,2$) is given by
\begin{equation}
\mathbf{F}_K(\mathbf{x}_1,\mathbf{x}_2)=C_0(f^0_{ph}+f^1_{ph}\ss_1\cdot\ss_2)\tt_1\cdot\tt_2
\delta(\mathbf{r}_1-\mathbf{r}_2),
\label{Fph}
\end{equation}
where $f^0_{ph}$ and $f^1_{ph}$ are the phenomenological Landau-Migdal
parameters. Hereafter, all the strength parameters of the residual interactions are given in units of $C_0=300$ MeV$\cdot$\,fm$^3$.

The residual p-p interaction $\mathbf{F}_K$ (for $K=3,4$) is given by a similar expression:
\begin{equation}
\mathbf{F}_K(\mathbf{x}_1,\mathbf{x}_2)=-C_0(g^0_{pp}+g^1_{pp}\ss_1\cdot\ss_2)
\delta(\mathbf{r}_1-\mathbf{r}_2),
\label{Gpp}
\end{equation}
and the pairing interaction:
\begin{equation}
\mathbf{F}^{pair}(\mathbf{x}_1,\mathbf{x}_2)=-C_0 g^{pair} \delta(\mathbf{r}_1-\mathbf{r}_2).
\label{Gpair}
\end{equation}
The pairing strengths $g^{pair}_n$ and $g^{pair}_p$ for neutron and proton subsystems are fixed within the BCS model 
to reproduce the experimental neutron and proton pairing energies. All the other strength parameters 
in the particle-particle channel are always given relative to $(g^{pair}_n+g^{pair}_p)/2$.

\subsection{Taking the single-particle continuum into consideration \label{cQRPA}}

The coordinate-space version of the pn-QRPA described in the preceding section 
is especially suitable for taking the s.p. continuum into consideration.
But before proceeding with the continuum, it is worth noting that
if one lets the double sums in Eq.~(\ref{A}) run just over finite sets of
proton and neutron s.p. states, 
the presented version of the pn-QRPA is fully equivalent to the usual ``discrete", one, 
which is formulated in terms of $X$ and $Y$ amplitudes. 
We make use of this fact in order to check the calculation scheme
by comparing ``discrete" QRPA results calculated in these two different, but formally equivalent, ways.
As anticipated, the results are the same within the accuracy of the numerical techniques used.

To take the s.p. continuum into consideration, the double-sum representation for the free response function (\ref{A})  should be transformed according to the following prescription: 
\begin{enumerate}

\item The Bogolyubov coefficients  $v_\alpha,\ u_\alpha$ and the quasiparticle energies $E_\alpha$ 
are approximated by their no-pairing values $v_\alpha=0,\ u_\alpha=1$, and $E_\alpha=\varepsilon_\alpha-\lambda_i$ for those s.p. states in the s.p. continuum that lie far up of the chemical potential $\lambda_i$ 
[i.e. $\varepsilon_\alpha-\lambda_i\gg \Delta_\alpha$, ``$i$"$=$ ``$p$" (protons) or ``$n$" (neutrons)]. 
The accuracy of this approximation is 
$\frac{\Delta}{|\varepsilon-\lambda|}$ which is good enough already for $\varepsilon_\alpha-\lambda_i \ge E_{max}\simeq$ 10 MeV.
The usual BCS representations for $v_\alpha,\ u_\alpha$ and $E_\alpha$ 
are taken for all the other s.p. states with $\varepsilon_\alpha-\lambda_i < E_{max}$. 

\item The radial single-particle Green's function 
$g_{(\alpha)}(r_1r_2,\varepsilon) =\displaystyle\sum_{\alpha_r}\frac{\chi_\alpha(r_1)\chi_\alpha(r_2)}
{\varepsilon-\varepsilon_\alpha+i\delta}$ is used to explicitly perform the sum over the s.p. states in the continuum.
Here, the sum $\sum_{\alpha_r}$ runs over different radial quantum numbers for a given spin-angular symmetry $(\alpha)$.
The Green's function satisfies the inhomogeneous radial s.p. Schr\"odinger equation 
$[h_{0(\alpha)}(r)-\varepsilon]g_{(\alpha)}(rr',\varepsilon)=-\delta(r-r')$
and can be constructed as a product of regular and irregular solutions of the homogeneous 
equation $[h_{0(\alpha)}(r)-\varepsilon]\chi_{(\alpha)}^{reg,irreg}(r,\varepsilon)=0$ 
(see, e.g., Refs.~\cite{Shlomo75,RingSchuck80}).
\end{enumerate}


As a result, we get from Eq. (\ref{A}) the following representation for the components $A_{IK}$ of the
free two-quasiparticle propagator: 
\bea
A_{11}(r_1r_2,\omega)&=&\sum\limits_{\nu_<,\pi_<}
\frac{v^2_\nu u^2_\pi}{\omega-E_{\pi\nu}}\chi_{\pi\nu}(r_1)\chi_{\pi\nu}(r_2)
+\sum\limits_{\nu_<,(\pi)}\left(t^{(JLS)}_{(\pi)(\nu)}\right )^2 v^2_\nu \,
\chi_\nu(r_1)\chi_\nu(r_2)\,g'_{(\pi)}(r_1r_2,\lambda_p+\omega-E_\nu)\nonumber\\
&&+ \left \{\pi\leftrightarrow\nu, \omega\to -\omega \right \}\label{propcont}\\
A_{12}(r_1r_2,\omega)&=&\sum\limits_{\nu_<,\pi_<}
u_\nu v_\nu u_\pi v_\pi\left[\frac{1}{\omega-E_{\pi\nu}}+\frac{1}{-\omega-E_{\pi\nu}}\right]
\chi_{\pi\nu}(r_1)\chi_{\pi\nu}(r_2)
\nonumber\\
A_{13}(r_1r_2,\omega)&=& \sum\limits_{\nu_<,\pi_<}u_\nu v_\nu \left[\frac{ u^2_\pi}{\omega-E_{\pi\nu}}
+\frac{v^2_\pi}{-\omega-E_{\pi\nu}}\right]\, \chi_{\pi\nu}(r_1)\chi_{\pi\nu}(r_2)
\nonumber\\
&&+\sum\limits_{\nu_<,(\pi)}\left(t^{(J)}_{(\pi)(\nu)}\right )^2 u_\nu v_\nu \,
\chi_\nu(r_1)\chi_\nu(r_2) \,g'_{(\pi)}(r_1r_2,\lambda_p+\omega-E_\nu)
\nonumber\\
A_{33}(r_1r_2,\omega)&=& \sum\limits_{\nu_<,\pi_<}
\left[\frac{u^2_\nu u^2_\pi}{\omega-E_{\pi\nu}}+\frac{v^2_\nu v^2_\pi}{-\omega-E_{\pi\nu}}\right]
\chi_{\pi\nu}(r_1)\chi_{\pi\nu}(r_2)\nonumber\\
&&
+\sum\limits_{\nu_<,(\pi)}\left(t^{(J)}_{(\pi)(\nu)}\right )^2 u^2_\nu\,
\chi_\nu(r_1)\chi_\nu(r_2)\,g'_{(\pi)}(r_1r_2,\lambda_p+\omega-E_\nu)\nonumber\\
&&+\sum\limits_{\pi_<,(\nu)}\left(t^{(J)}_{(\pi)(\nu)}\right )^2 u^2_\pi\,
\chi_\pi(r_1)\chi_\pi(r_2)\,g'_{(\nu)}(r_1r_2,\lambda_n+\omega-E_\pi)
\nonumber\\
A_{44}(\omega)&=&A_{33}(-\omega);\ A_{14}(\omega)=A_{13}(-\omega,\pi\leftrightarrow\nu);\ A_{24}(\omega)=A_{13}(-\omega);\ 
A_{23}(\omega)=A_{13}(\omega,\pi\leftrightarrow\nu) \nonumber
\eea
where $\pi_{<}$ ($\nu_{<}$) means $\pi\le\pi_{max}$ ($\nu\le\nu_{max}$), where $\pi_{max}$ 
($\nu_{max}$) is the s.p. state with the largest energy included into the BCS basis for which 
$E_{\pi_{max}}=\varepsilon_{\pi_{max}}-\lambda_{p}$ ($E_{\nu_{max}}=\varepsilon_{\nu_{max}}-\lambda_{n}$) 
with the acceptable accuracy as previously described, and 
\be
g'_{(\pi)}(r_1r_2,\varepsilon)=g_{(\pi)}(r_1r_2,\varepsilon)-
\sum\limits_{\pi_{r\, <}}\frac{\chi_\pi(r_1)\chi_\pi(r_2)}{\varepsilon-\varepsilon_\pi}
\label{greenf}
\ee
is the subtracted radial s.p. Green's function (the Green's function from which the contribution 
of all discrete s.p. states and those quasidiscrete s.p. states included in the BCS basis 
is subtracted). 


\subsection{Strength functions\label{SF}}

Different strength functions can be readily calculated in terms of the imaginary part of the total two-quasiparticle propagator $\Image\tilde A$. 
The strength function corresponding to a charge-exchange single-particle operator $V^{(\mp)}_{J\mu}$ 
acting in the $\beta^{(\mp)}$ channel,
\be
\hat V^{(\mp)}_{JLSM}=\sum_a {\mathbf V}_{JLSM}({\mathbf x}_a)
\tau_a^{(\mp)},
\ee
where ${\mathbf V}_{JLSM}({\mathbf x}_a)$ is given by Eq.~(\ref{Vm}),
is defined by the usual expression:
\be
S^{(\mp)}(\omega)=\sum\limits_{s} \left | \langle J^\pi M, s | \hat V^{(\mp)}_{JLSM} |0\rangle\right |^2
\delta(\omega-\omega_s^\mp)\nonumber
\ee
with $\omega_s^\mp=E_s^\mp-E_0$ being the excitation energy of the corresponding
isobaric nucleus $(N\mp 1,Z\pm 1)$ 
relative to the ground state $|0\rangle$ of the parent nucleus $(N,Z)$ with energy $E_0$.
Making use of the spectral decomposition (\ref{spectrA}) one can easily verify the following
integral representations of the strength functions:
\bea
&S^{(-)}(\omega^-)=-\frac 1\pi \left\{{\mathbf V} \mathbf{ \tilde A}_{11}(\omega) {\mathbf V} \right\}=-\frac 1\pi \Image \int V_{(JLS)}(r_1) \tilde A_{11}^{(JLS)}(r_1r_2;\omega) V_{(JLS)}(r_2)\, dr_1dr_2, \label{Sm}\\
&S^{(+)}(\omega^+)=-\frac 1\pi \left\{{\mathbf V} \mathbf{\tilde A}_{22}(\omega) {\mathbf V} \right\}
=-\frac 1\pi \Image \int V_{(JLS)}(r_1) \tilde A_{22}^{(JLS)}(r_1r_2;\omega) V_{(JLS)}(r_2)\, dr_1dr_2. \label{Sp}
\eea
where $\omega^\mp=\omega\pm (\lambda_p-\lambda_n)\pm (m_n-m_p)$ represents the calculated excitation energy relative to 
the g.s. of the parent nucleus and $m_p$, $m_n$ are proton and neutron masses.
The pn-QRPA excitation spectrum, originally calculated in terms of $\omega$, gets a constant energy shift 
to be represented in terms of $\omega^\mp$,
because the modified nuclear Hamiltonian $\hat H - \lambda_p\hat Z - \lambda_n\hat N$ 
(as in the BCS model) is used within the QRPA 
and the model nuclear Hamiltonian does not contain the rest energies of nucleons.

One can also define a nondiagonal strength function as 
\be
S^{(- -)}_V(\omega)=\sum\limits_{s}  \langle 0' | \hat V^{(-)}_{JLS\bar M} |J^\pi M, s \rangle 
\langle J^\pi M, s | \hat V^{(-)}_{JLSM} |0\rangle \delta(\omega -\omega'_s)
\ee
with $\bar\omega_s=E_s-(E_0+E_{0'})/2=(\omega_s^-+{\omega'}_s^+)/2$ and $\hat V^{(-)}_{JLS\bar M}$ being 
the time reverse of $\hat V^{(-)}_{JLSM}$. Such a strength function is closely 
related to the amplitude of the \bb \ decay, when $|0\rangle$ and $|0'\rangle$ are the g.s. wave functions of the initial (decaying) and final (product) nuclei, respectively.
 
To calculate $S^{(- -)}_V(\omega)$
within the pn-QRPA one faces the usual problem that the spectrum $|s\rangle$ comes out
slightly different when calculated with respect to $|0\rangle$ or $|0'\rangle$.
Identifying the QRPA vacuum $|0'\rangle$ with that of $|0\rangle$, one gets $\bar\omega_s=\omega_s$ and
\be
S^{(- -)}(\omega)=-\frac 1\pi \left\{{\mathbf V} \mathbf{\tilde A}_{12}(\omega) {\mathbf V}\right\}=
-\frac 1\pi \Image\int V_{(JLS)}(r_1) \tilde A_{12}^{(JLS)}(r_1r_2;\omega) V_{(JLS)}(r_2) \, dr_1dr_2
\label{Smm}
\ee
or, alternatively, identifying $|0\rangle$ with $|0'\rangle$
\be
S^{(- -)}(\omega)=-\frac 1\pi \left\{{\mathbf V} \mathbf{\tilde A'}_{12}(\omega) {\mathbf V}\right\}=
-\frac 1\pi \Image\int V_{(JLS)}(r_1)\tilde {A'}^{(JLS)}_{12}(r_1r_2;\omega) V_{(JLS)}(r_2) \, dr_1dr_2
\ee
where $	\tilde A'$ is calculated with respect to the g.s. $|0'\rangle$ of the final nucleus.

\subsection{Description of $\beta\beta$ decay within the cQRPA \label{bb}}

The spectral decomposition of the two-quasiparticle propagator $\tilde A$ (\ref{spectrA}) 
can be used for calculation of $\beta\beta$ decay matrix elements in a similar way as described in
Sec.~\ref{SF} for $S^{(- -)}(\omega)$ (\ref{Smm}). 

The \bb \ decay amplitude $M^{2\nu}_{GT}$ is defined by the following expression:
\be
M^{2\nu}_{GT}=\sum\limits_{s} \frac{ \langle 0' \| \hat G^{(-)} \|s \rangle 
\langle s \| \hat G^{(-)} \|0\rangle } {\bar\omega_s}
\label{defM2nu}
\ee
where $\hat G^{(-)}=\sum_a \ss_a\ \tau^-_a$ and again $\bar\omega_s=E_s-(E_0+E_{0'})/2=(\omega_s^-+{\omega'}_s^+)/2$. 

By using the spectral decomposition of Eqs.~(\ref{spectrA}) for $(JLS)=(101)$ and the approximation that the QRPA vacuum 
$|0'\rangle$ of the final g.s. is the same as $|0\rangle$ of the initial g.s. (the same approximation 
as used in Refs.~\cite{vog86,Engel88}), the amplitude (\ref{defM2nu}) is simply given by one-half of the corresponding static nuclear polarizability with respect to the external s.p. field $\ss\tau^-$:
\bea
&&M^{2\nu}_{GT}=-\frac 12 \left\{\ss \mathbf{\tilde A}_{12}(\omega=0)  \ss\right\}=-6\pi \int\tilde A^{(101)}_{12}(r_1r_2;\omega=0) dr_1dr_2\label{M2nu1}
\nonumber
\eea
where $\bar\omega_s=\omega_s$ is used consistently in this approximation.

The same procedure can be applied to calculate within the cQRPA the matrix element $\langle 0' | \hat W^{(- -)} |0\rangle$ of a two-body scalar operator
\bea
\displaystyle\hat W^{(- -)}=\sum_{ab} \mathbf{W}(\mathbf{x}_a,\mathbf{x}_b)\tau_a^{(-)}\tau_b^{(-)}\label{V2}\\
\mathbf{W}(\mathbf{x}_a,\mathbf{x}_b)=\sum_{JLSM} W_{(JLS)}(r_a,r_b)T_{JLSM}(\mathbf{ n}_a)T^*_{JLSM}(\mathbf{ n}_b)\nonumber
\eea
between the ground states $|0\rangle$ and $|0'\rangle$. It is given by a sum of all partial contributions $M^{(JLS)}$:
\bea
&M^{(- -)} 
=-\frac 1\pi\int d\omega \, \Image\left\{\mathbf{W} \mathbf{\tilde A}_{12}(\omega) \right\}=\displaystyle\sum_{JL} M^{(JLS)}\label{M--}\\
&M^{(JLS)} = -\displaystyle\frac{(2J+1)}{\pi}  \int d\omega \int W_{(JLS)}(r_1,r_2) \Image \tilde A^{(JLS)}_{12}(r_1r_2;\omega) \, dr_1dr_2
\label{M---}
\eea
(where the identification of the ground states as described previously has to be done).

The neutrino potential appearing in the description of the \bbzn \ decay (see,e.g., Refs.~\cite{fae98,Suh98}):

\be
\displaystyle\hat W^{(- -)}_{0\nu}=\sum_{ab} P_\nu(r_{ab}) \left(\ss_a \cdot \ss_b -\frac{g_V^2}{g_A^2}\right)
\tau_a^{-} \tau_b^{-}
\label{nuPot}
\ee
in the simplest 
Coulomb approximation $P_\nu(|\vec r_a-\vec r_b|\equiv r_{ab})=R/r_{ab}$ (with $R=1.23\,A^{1/3}$ fm 
being the nuclear radius) has the well-known partial radial components
$W_{(JLS)}(r_a,r_b)=\frac{4\pi}{2L+1}
\frac{R}{r_>}\left(\frac{r_<}{r_>}\right)^L$ 
(where $r_<=\mathrm{min}(r_a,r_b),\ r_>=\mathrm{max}(r_a,r_b)$).
When one takes into account both the energy dependence of the neutrino 
potential $P_\nu=P_\nu(r_{ab},\omega)$ and the usual 
modification $P_\nu\to P_\nu f_J^2 $ with the Jastrow-like function $f_J(r_{ab}) = 1 - e^{-\gamma_1 r_{ab}^2}(1 - \gamma_2 r_{ab}^2)$ to account for the SRC of the two initial neutrons and two final protons 
, the decomposition of the neutrino potential over the Legendre polynomials $P_L$ 
can be done numerically:
\bea
& W^{0\nu}_{(JLS)}(r_1,r_2,\omega) = \displaystyle\frac{(2L+1)}{2} \int\limits_0^\pi d 
\cos\theta_{12}
\, P_L(
\cos\theta_{12})\, P_\nu(r_{12},\omega).
\label{Leg}
\eea

In this first application of the cQRPA the corrections from the high-order terms in the nucleonic weak current 
and the finite nucleon size are omitted, which can lead to a slight overestimation of the calculated $M^{0\nu}$ 
by 20--30 \%~\cite{Rod05}.

Note that, within the cQRPA, in contrast to the discrete QRPA, one does not get an explicit set of QRPA energies and the energy integrations  in the expressions for $M^{0\nu}$ (\ref{M--}),(\ref{M---}) have to be performed on a mesh. For each  point in the energy mesh the cQRPA equations (\ref{Atilde}) are solved by discretizing the spatial integrals, thereby transforming them to a matrix representation. 
All this makes the calculation of $M^{2\nu}$ and $M^{0\nu}$ rather time consuming. 
Implementation of adaptive integration methods helps to optimize the integration over the energy.

\section{Calculation results \label{results}}

We perform the first calculations of the $\beta\beta$ transition amplitudes $M^{2\nu}$ and $M^{0\nu}$ within the cQRPA for $^{76}$Ge, $^{100}$Mo and $^{130}$Te.
We also compare the results obtained with those calculated within the usual ``discrete"
version of the QRPA to see the influence of the single-particle continuum.

For the first calculations of $M^{2\nu}$ and $M^{0\nu}$ within the continuum QRPA we adopt
a rather simple nuclear Hamiltonian similar to that used in Refs.~\cite{vog86,Engel88}.
The chosen nuclear mean field $U(x)$ consists of the
phenomenological isoscalar part $U_0(x)$ along with the isovector $U_1(x)$
and the Coulomb $U_C(x)$ parts, both calculated consistently in
the Hartree approximation (see Ref.~\cite{Rod03}). 
The residual isovector particle-hole interaction 
and the particle-particle interaction in both the neutral (pairing) and charge-exchange channels are chosen in the form of 
zero-range forces~[Eqs~(\ref{Fph})-\ref{Gpair})]. 
All the strength parameters of the residual interactions are given 
in units of 300 MeV$\cdot$\,fm$^3$.

The results calculated within the discrete-QRPA, labeled A, B, C, refer to different s.p. bases used.
Case  A corresponds to the large s.p. basis in the calculations: 16 successive s.p. levels comprising $N=1-5$ major Saxon-Woods shells for $^{76}$Ge and $^{100}$Mo and 22 successive s.p. levels (all bound s.p. states for neutrons
and all bound s.p. states along with 6 quasistationary states for protons) comprising $N=1-6$ major shells for $^{130}$Te. Note that the same s.p. basis is used within the cQRPA as the basis of the BCS problem. 
Case B corresponds to the 
small s.p. basis and is obtained from A by subtracting the six lowest s.p. levels comprising $N=1-3$ major shells 
(inert core of $^{40}$Ca).
Case  C corresponds to the solution of the QRPA equations in the large basis A, in which, however, 
the BCS problem is solved in the small basis B (i.e. the Bogolyubov coefficients 
$v_\alpha=1$ are taken for the six lowest s.p. levels).
This tests the approximations involved in the current cQRPA calculations, namely, 
that $v_\alpha$ is set exactly to zero for the s.p. states lying above the BCS basis.

Fixing the model parameters is done in the following way: 
\begin{itemize}
\item The p-h isovector strength $f^0_{ph}$ is chosen equal to unity, $f^0_{ph}=1.0$.
This allows reproduction of the experimental nucleon binding energies for closed-shell nuclei
by implementing the isospin self-consistency of the symmetry potential 
$U_1(x)$ of the mean field with the isovector p-h interaction (see Ref.~\cite{Rod03}).
\item The p-h spin-isovector strength $f^1_{ph}$ is fitted to reproduce the experimental
energy of the GTR.
\item The pairing strengths $g^{pair}_n$, $g^{pair}_p$ are fixed within the BCS model 
to reproduce the experimental pairing energies. As already mentioned, all the other 
strength parameters in the particle-particle channel are given relative to $(g^{pair}_n+g^{pair}_p)/2$.
\item  By choosing the p-p isovector strength $g^0_{pp}=1.0$ we restore
approximately the isospin self-consistency of the total residual p-p interaction.
\item The p-p spin-isovector strength $g^1_{pp}$ is chosen to reproduce the experimental (positive) value of $M^{2\nu}$
as done in Refs.~\cite{Rod03a,Rod05}.
\end{itemize}

In Fig.~1 the calculated $g^1_{pp}$ dependence of $M^{2\nu}$ is plotted. Calculated $M^{2\nu}$ depends on the choice of the g.s. (either initial or final) with respect to which the QRPA equations are solved.
It can be seen in the figure that the $M^{2\nu}$ calculated within the discrete QRPA and the cQRPA are almost the same 
at small values of $g^1_{pp}$, though the difference becomes visible while $g^1_{pp}$ grows.
Thus, the correction to the $M^{2\nu}$ coming from the s.p. continuum is small, 
as can be expected for the Gamow-Teller transitions.

\begin{figure}[tb]
\label{fig2nu}
\begin{center}
\includegraphics[scale=0.75]{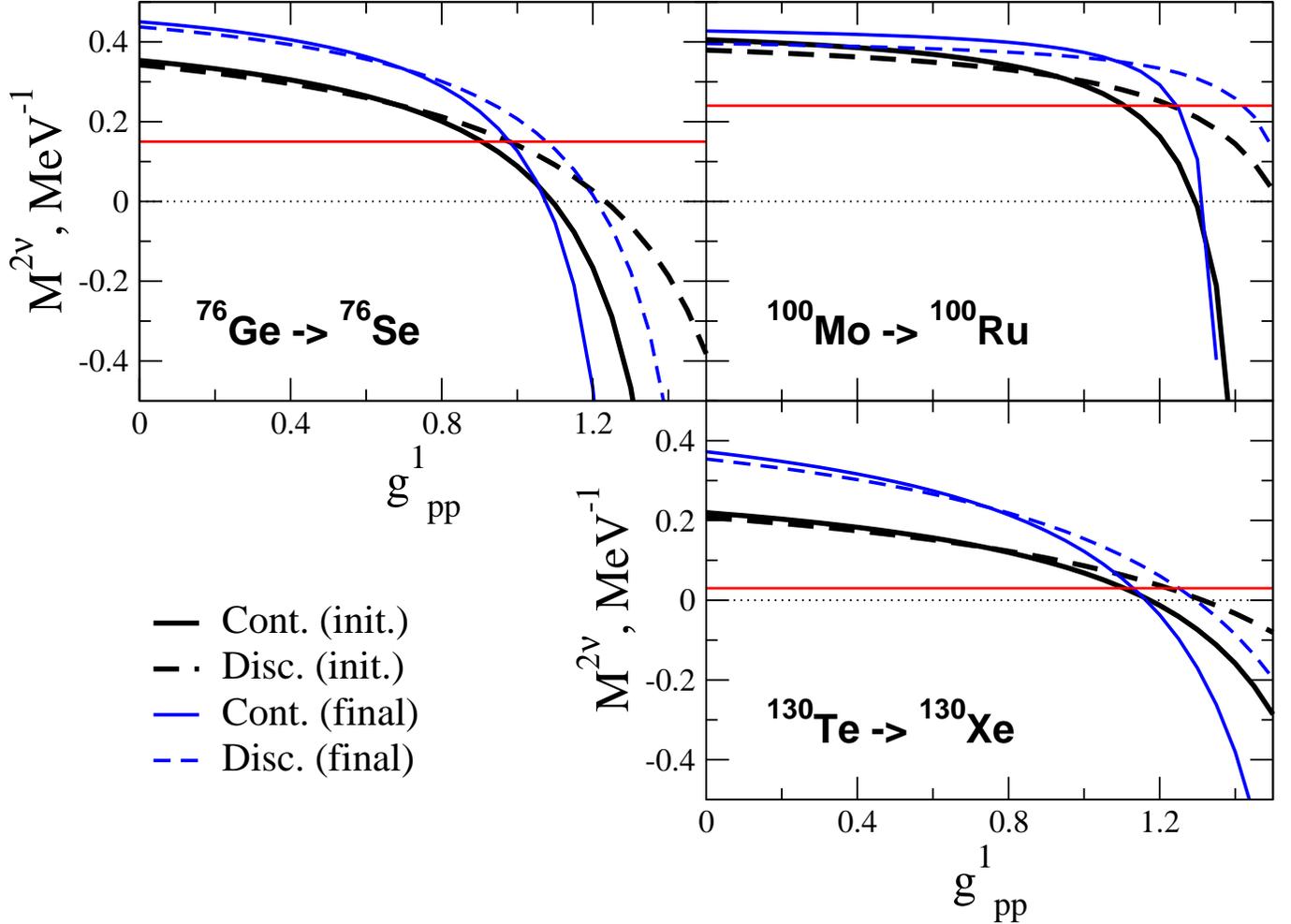}
\caption{(Color online) The calculated dependence of the $2\nu\beta\beta$ matrix element on the p-p strength $g^1_{pp}$
in both discrete (set A, dashed lines) and continuum QRPA (solid lines).
The calculations performed by using the QRPA set of the intermediate states built with respect to the g.s of the initial nucleus(e.g. $^{76}$Ge) and the g.s. of the final nucleus (e.g. $^{76}$Se) 
are depicted by thick and thin lines, respectively. 
The solid horizontal line gives the (positive) experimental value of the $2\nu\beta\beta$ decay matrix elements 
used in the calculation to fit $g^1_{pp}$.}
\end{center}
\end{figure}

The values of the strength parameters $f^1_{ph}$ and $g^1_{pp}$ fixed according to our prescription are listed in
Table~\ref{tab1} for both the discrete and continuum versions of the QRPA. Because calculated $M^{2\nu}$ depends on the choice of the g.s. (either initial or final) with respect to which the QRPA equations are solved, two sets of $g^1_{pp}$ are obtained. The upper and lower lines for each decay sequence in Table~\ref{tab1} contain $g^1_{pp}$ fitted for initial and final nuclei, respectively. One sees that the difference in the obtained $g^1_{pp}$ is almost negligible for $^{130}$Te$\to ^{130}$Xe, but it becomes $\delta g^1_{pp}\approx 0.2$ for $^{100}$Mo$\to ^{100}$Ru.

\begin{table}[t] 
  \begin{center} 
    \caption{Choice of the model parameters. See text for explanation 
of the choices A, B, C for different model spaces within the discrete QRPA.
The two lines for the parameters of $g^1_{pp}$  for each $\beta\beta$ decay are fitted  
by using the QRPA set of the intermediate states built with respect 
to the g.s of the initial nucleus (e.g. $^{76}$Ge) and the g.s. of the final nucleus (e.g. $^{76}$Se).
The p-h strength $f^1_{pp}$ is adjusted to reproduce the energy of the Gamow-Teller resonance in 
the initial nucleus (e.g. $^{76}$Ge).} 
\label{tab1} 
\begin{tabular}{|l|c|c|ccc||c|} 
\hline\hline 
 Nuclear& $M^{2\nu}_{exp}$, & strength & \multicolumn{3}{|c||} {\ discrete QRPA \ } & {continuum} 
\\ \cline{4-6} 
transition& MeV$^{-1}$ & parameters & \phantom{Nu}A\phantom{Nu} & \phantom{Nu}B\phantom{Nu} & \phantom{Nu}C\phantom{Nu} & QRPA\\

\hline  

 & &$g^1_{pp}$ & 0.98 & 0.97 & 0.97 & 0.91\\
$^{76}$Ge$\to^{76}$Se & 0.15 & $g^1_{pp}$ & 1.10 & 1.10 & 1.10 & 1.01\\
          & & $f^1_{ph}$ & 0.40 & 0.43 & 0.43 & 0.40 \\ 
\hline
         &  & $g^1_{pp}$ & 1.28 & 1.31 & 1.30 & 1.10\\
$^{100}$Mo$\to^{100}$Ru     & 0.24 & $g^1_{pp}$ & 1.43 & 1.50 & 1.49 & 1.23\\
          &  &$f^1_{ph}$ & 0.70 & 0.70 & 0.70 & 0.75 \\
\hline 
 &  & $g^1_{pp}$ & 1.23 & 1.25 & 1.25 & 1.11\\
$^{130}$Te$\to^{130}$Xe & 0.03 & $g^1_{pp}$ & 1.25 & 1.28 & 1.28 & 1.13\\   
          &  & $f^1_{ph}$ & 0.60 & 0.60 & 0.60 & 0.63 \\ 
\hline\hline 
\end{tabular} 
  \end{center} 
\end{table}

The calculated values of $M^{0\nu}$ are given in Table~\ref{tab2} for both versions of the QRPA for $g_A=1.25$.
The numbers in parentheses are the matrix elements calculated with inclusion of the SRC in terms of the Jastrow-like function. 
The two lines of results for each $\beta\beta$ decay chain contain $M^{0\nu}$ calculated with respect to 
the g.s. of the initial or final nucleus in the decay (see Sec.~\ref{bb}). 
In the present first application of the cQRPA, 
neither the finite nucleon size nor the higher order terms of the nucleon current are considered. 
(They usually bring an additional reduction of $M^{0\nu}$ by about 30\%; see, e.g., Ref.~\cite{Rod05}). 

The contribution of the multipoles with $L=0-5$ are included in the calculations of $M^{0\nu}$. 
The contributions with $L>5$ (which can increase $M^{0\nu}$ in total by about 10\%) are omitted
here as the corresponding parts in the transition operator probe the short-range 
behavior of the nucleon-nucleon wave function that cannot be well described within the QRPA. 
It is known that the RPA in medium is formulated to describe 
propagation of small-amplitude density fluctuations and
only the ring diagrams are summed (see, e.g., Ref.~\cite{FetWal71}).
This is a quite suitable approximation to deal with 
collective long-wave excitations, but for the short-range ones
the diagrams that are left out of the RPA method become important.

As one sees from comparison of the discrete QRPA results listed in columns A and B, 
the calculated $M^{0\nu}$ values for different basis sizes come out very close to each other provided 
the p-p interaction parameter $g^1_{pp}$ is fixed to reproduce the experimental \bb \ decay matrix elements
$M^{2\nu}_{exp}$. This result provides an independent confirmation of the main conclusion of Ref.~\cite{Rod05} which is 
obtained here for a different nuclear Hamiltonian and by solving the pn-QRPA equations in the coordinate representation.
Also, the $M^{0\nu}$ values 
obtained by using only initial QRPA g.s. (upper line) or final one (lower line) in the calculation 
are rather closer to each other.

The matrix elements $M^{0\nu}$ calculated by taking into account the SRC in terms of the Jastrow functions
get suppressed by about 25--30 \% in the present calculation, that is in quantitative agreement with other 
recent calculations~\cite{Rod05,Suh07a} but do not support the old results of Ref.~\cite{Engel88} where the 
strong suppression was found. 
However, one should keep in mind that this way of describing the SRC can be rather rough and 
can lead to overestimation of the suppression of the $M^{0\nu}$. 
Other methods of describing the SRC such as UCOM~\cite{UCOM} give much less suppression in 
the calculated $M^{0\nu}$ (only about 10 \%) and this important issue is currently under 
intensive study~\cite{Suh07a,Sim07}.

The matrix elements $M^{0\nu}$ calculated within the cQRPA (the last column of Table~\ref{tab2}), they are systematically smaller than the discrete QRPA ones (columns A and B). The suppression varies from about 30 \% for $^{76}$Ge to a factor 2 for $^{100}$Mo and $^{130}$Te. The origin of this suppression can be associated 
with additional ground state correlations appearing because of 
highly excited collective states embedded in the s.p. continuum. 
Transitions to these states are naturally described within the cQRPA 
in terms of the s.p. Green's functions (see Sec.~\ref{cQRPA}).
However, the applied version of the cQRPA in this work does not include nucleon pairing in the s.p. continuum and
therefore the $M^{0\nu}$ values obtained here should be treated only as lower limits. 
Inclusion of nucleon pairing in the s.p. continuum (which is a formidable task) 
will definitely lead to an increase of the matrix elements 
within the cQRPA.

To demonstrate the importance of nucleon pairing far from the Fermi level for quantitative description of the $M^{0\nu}$,
the numbers listed in the column C of Table~\ref{tab2} can be compared with those in columns A and B. Case C is introduced, as previously described, to test within the discrete-QRPA the neglect of pairing far from the Fermi level, in a manner similar to how it is done in the present version of the cQRPA. Inspecting column C, one sees a marked reduction, by about 30 \%, in the calculated \bbzn \ matrix elements. Thus, expanding the 
discrete QRPA basis from the ``small" one of B to a ``large" one of C which neglects pairing effects in the inert core
but allows transitions from the inert core, leads to a suppression in the $M^{0\nu}$ because of more g.s. correlations. The suppression, however, gets almost completely compensated as nucleon pairing is switched on in the inert core and one goes from case C to case A. The same sort of compensation is natural to expect in the case of the cQRPA when nucleon pairing is switched on in the single-particle continuum. However, one cannot exclude that 
the compensation is incomplete. 

Let us conclude with some words about possible prospects for taking 
nucleon pairing in the s.p. continuum into consideration within the approach described in here.
Though possible ways of treating the continuum pairing within the QRPA can be found in the literature (see, e.g., 
Ref.~\cite{Mats01}), direct implementation of them would drastically increase the corresponding calculation efforts.
One would need first to calculate the solutions $u(r)$ and $v(r)$ of the coordinate Hartree-Fock-Bogolyubov equation
for positive energies, from which then additional continuum contributions to the expressions (\ref{propcont}) for the response function should be constructed by direct integration  over energy. 
Probably, a more economical way is to discretize the continuum by putting the nucleus in a large box. 
These further developments are postponed to a future publication that should then finally answer the question 
about stability of $M^{0\nu}$ with respect to the basis size.

\begin{table}[t] 
  \begin{center} 
\caption{$0\nu\beta\beta$ nuclear matrix elements evaluated without and with (in parentheses) the SRC
in both discrete and continuum QRPA. See text for explanation 
of the choices A, B, C for different model spaces within the discrete QRPA.
The two lines of results for each $\beta\beta$ decay contain $M^{0\nu}$ 
calculated by using the QRPA set of the intermediate states built with respect 
to the g.s of the initial nucleus (e.g. $^{76}$Ge) and the g.s. of the final nucleus (e.g. $^{76}$Se).} 
\label{tab2} 
\begin{tabular}{|l|c|c|c||c|} 
\hline\hline 
 Nuclear &  \multicolumn{3}{|c||} {\ discrete QRPA \ } & {continuum} 
\\ \cline{2-4} 
transition &  \phantom{Nu}A\phantom{Nu} & \phantom{Nu}B\phantom{Nu} & \phantom{Nu}C\phantom{Nu} & QRPA\\
\hline  
$^{76}$Ge & 5.95 (4.30) & 5.63 (4.19) & 4.30 (3.19) & 4.30 (3.09)\\
$\to\ ^{76}$Se & 5.44 (3.86)  & 5.22 (3.82) & 3.81 (2.76) & 3.63 (2.46) \\
\hline
$^{100}$Mo & 5.52 (3.88) & 5.35 (3.84) & 4.24 (3.00)& 2.49 (1.67)\\
$\to\ ^{100}$Ru & 4.19 (2.73) & 4.00 (2.65) & 2.91 (1.84)& 1.39 (0.67)\\    
\hline 
$^{130}$Te & 3.17 (2.19) & 3.14 (2.20) & 2.56 (1.78) & 1.70 (1.12) \\ 
$\to\ ^{130}$Xe & 4.69 (3.21) & 4.67 (3.21) & 3.77 (2.57) & 2.03 (1.28)\\
\hline\hline 
\end{tabular} 
 \end{center} 
\end{table} 

\section{Conclusions \label{conclusions}}

A continuum-QRPA approach to calculation of the nuclear double beta decay
\bb- and \bbzn-amplitudes has been formulated. Calculations of the amplitudes $M^{2\nu}$ and $M^{0\nu}$
within the cQRPA are performed for $^{76}$Ge, $^{100}$Mo and $^{130}$Te.
A rather simple nuclear Hamiltonian consisting of phenomenological mean field 
and zero-range residual particle-hole and particle-particle interaction is used.
The $M^{2\nu}$ are almost unaffected when the single-particle continuum is taken into account. 
In contrast, we find a regular suppression of the \bbzn \ amplitude
that can be associated with additional ground state correlations 
owing to collective states in the continuum. 
The calculated $M^{0\nu}$ values of this paper should be considered as lower limits 
for the matrix elements within the cQRPA, as nucleon pairing is realized 
only on a discrete basis within the present version of the cQRPA.
It is expected that future inclusion
of nucleon pairing in the single-particle continuum will somewhat 
compensate the observed suppression of $M^{0\nu}$ values.

\subsection*{\it Acknowledgments} 
We are grateful to  Mrs. L. Rodina for helping us with 
implementation of an adaptive integration procedure in the numerical solution 
of the continuum-QRPA equations. We also thank Prof. M.~Urin for valuable discussions.
The work is supported in part by the Deutsche Forschungsgemeinschaft 
(by both grant FA67/28-2 and the Transregio Project TR27 ``Neutrinos and Beyond") 
and by the EU ILIAS project (Contract RII3-CT-2004-506222).

\end{document}